\documentclass[aps,prb,twocolumn,superscriptaddress]{revtex4-2}

\usepackage{pifont}
\usepackage[colorlinks=true,citecolor=blue,urlcolor=blue,linkcolor=blue,hyperfigures=true]{hyperref}
\usepackage{graphicx}
\usepackage{amsmath,amsfonts,amssymb}
\usepackage[table,x11names]{xcolor}
\usepackage{color}
\usepackage{changes}
\usepackage{booktabs}
\usepackage{ulem}
\DeclareUnicodeCharacter{2212}{ }

\newcommand{\iw}{i\omega}
\newcommand{\inu}{i\nu}

\begin{document}

\title{Screening Induced Crossover between Phonon- and Plasmon-Mediated Pairing in Layered Superconductors}

\author{Y. in 't Veld}
\affiliation{Institute for Molecules and Materials, Radboud University, 6525 AJ Nijmegen, the Netherlands}
\author{M.I. Katsnelson}
\affiliation{Institute for Molecules and Materials, Radboud University, 6525 AJ Nijmegen, the Netherlands}
\author{A.J. Millis}
\affiliation{Center for Computational Quantum Physics, Flatiron Institute, New York, NY 10010, United States of America}
\affiliation{Department of Physics, Columbia University, New York, NY 10027, United States of America}
\author{M. R\"osner}
\affiliation{Institute for Molecules and Materials, Radboud University, 6525 AJ Nijmegen, the Netherlands}

\date{\today}

\begin{abstract}
  Two-dimensional (2D) metals can host gapless plasmonic excitations, which strongly couple to electrons and thus may significantly affect superconductivity in layered materials. To investigate the dynamical interplay of the electron-electron and electron-phonon interactions in the theory of 2D superconductivity, we apply a full momentum- and frequency-dependent one-loop theory treating electron-phonon, electron-plasmon, and phonon-plasmon coupling with the same accuracy. We tune the strength of the Coulomb interaction by varying the external screening $\varepsilon_{ext}$ to the layered superconductor and find three distinct regions. At weak screening, superconductivity is mediated by plasmons. In the opposite limit conventional electron-phonon interactions dominate. In between, we find a suppressed superconducting state. Our results show that even conventional electron-phonon mediated layered superconductors can be significantly affected by the electron-plasmon coupling in a weak screening environment. This unconventional pairing contribution can then be controlled by the external screening. 
\end{abstract}

\maketitle

\textbf{Introduction.} Superconductivity in ultra-thin two-dimensional (2D) films is of long-standing scientific importance and has undergone a recent revival of interest. 
Experimental studies of atomically thin elemental superconductors~\cite{qin_superconductivity_2009, zhang_superconductivity_2010,van_weerdenburg_threefold_2022}, superconducting 2D electron gases formed in layered oxide heterostructures~\cite{ohtomo_high-mobility_2004,reyren_superconducting_2007}, as well as 2D van-der-Waals materials such as transition metal dichalcogenides~\cite{frindt_superconductivity_1972,ye_superconducting_2012,cao_quality_2015,costanzo_gate-induced_2016,xi_ising_2016,khestanova_unusual_2018,wang_high-quality_2017}, FeSe~\cite{wang_interface-induced_2012,zhang_effects_2016,zhou_dipolar_2017,zhang_enhanced_2019,wei_high-temperature_2021}, and various forms of multilayer graphene~\cite{kanetani_ca_2012,margine_electron-phonon_2016,chapman_superconductivity_2016,wang_strong_2022} on various substrates have produced results that seem to challenge the conventional understanding of superconductivity. Most recently, the ability to tune properties of ultra-thin materials by gate voltages and novel heterostructuring including ``moir\'e'' systems~\cite{andersen_dielectric_2015,schonhoff_interplay_2016,cao_unconventional_2018,hall_environmental_2019,steinke_coulomb-engineered_2020,da_jornada_universal_2020,jiang_plasmonic_2021} has further increased the interest in 2D superconductivity and rendered layered superconductors a promising platform for many-body material design.

The origin of superconductivity observed in atomically thin 2D materials remains debated. In some materials unconventional coupling schemes due to charge~\cite{frohlich_superconductivity_1968, frindt_superconductivity_1972,takada_plasmon_1978,takada_plasmon_1992,bill_electronic_2003,akashi_development_2013,akashi_density_2014,sano_plasmon_2019,sanna_combining_2020,wei_high-temperature_2021,davydov_ab_2020,akashi_revisiting_2022,cai_superconductivity_2022,grankin_interplay_2022} or spin~\cite{mazin_unconventional_2008, yin_spin_2014, acharya_electronic_2021} fluctuations have been proposed, while for other compounds conventional electron-phonon coupling has been proposed as the origin of pairing ~\cite{calandra_effect_2009,ge_phonon-mediated_2013,li_electron-phonon_2014, rosner_phase_2014,coh_large_2015, zhang_ubiquitous_2017}.
Even the conventional theory of 2D superconductivity is challenging since in ultra-thin systems reduced environmental screening yields Coulomb interaction with enhanced long-range character in space and strong retardation in frequency, yielding gapless $\sqrt{q}$-like plasmon excitations with diverging electron-plasmon coupling in the long-wavelength limit~\cite{pines_approach_1962,caruso_two-dimensional_2021}. These particular plasmonic properties have a non-trivial influence on superconductivity that is qualitatively different from any phonon mediated mechanisms. Additionally, the low-energetic gapless plasmon dispersion allows for an efficient hybridization between plasmons and optical phonons~\cite{dong_coupled_2020}. Thus, in contrast to conventional 3D superconductors, in which plasmons have large gaps, a fully momentum-dependent and dynamical theory is required to accurately describe 2D superconductors in the presence of electron-phonon and electron-electron interactions.~\cite{eliashberg_interactions_1960,eliashberg_temperature_1960,akashi_development_2013,akashi_density_2014,sanna_combining_2020,davydov_ab_2020,akashi_revisiting_2022,simonato_revised_2022} 
Furthermore, even for 3D materials the commonly used Tolmachev-Morel-Anderson pseudopotential $\mu^*$ approximation for the Coulomb interaction~\cite{tolmachev_logarithmic_1961,morel_calculation_1962} has been called in to question~\cite{wang_origin_2022,akashi_revisiting_2022}.

Here we present a first step towards addressing these fundamental theoretical issues at hand of a generic monolayer model with some finite thickness $h$ and a single occupied electron pocket. Focusing on $s$-wave superconductivity initially mediated by dispersionless optical phonons with and without non-local electron-phonon coupling in the presence of the 2D Coulomb interaction, we present a consistent theory including both electron-electron and electron-phonon interactions on the same one-loop theoretical level. We study the interplay of electron-phonon, electron-plasmon, and phonon-plasmon interactions effects on normal state as well as superconducting properties as a function of the external screening and find that conventional electron-phonon mediated superconductors can be driven to an unconventional plasmon-mediated regime when the overall screening is small.

\textbf{Formalism.} We use a non-local 2D background-screened Coulomb interaction $v_C(q) = 2 \pi e^2 / (A \varepsilon_{back}(q) q)$ with $\varepsilon_{back}(q)$ being the background dielectric function given by,
\begin{equation}
    \varepsilon_{back}(q) = \varepsilon_{int} \frac
    {1 - \tilde{\varepsilon}^2 e^{-2 q h}}
    {1 + 2 \tilde{\varepsilon} e^{-qh} + \tilde{\varepsilon}^2 e^{-2 q h}},
\end{equation}
with $\tilde{\varepsilon} = \left(\varepsilon_{int} - \varepsilon_{ext}\right) / \left(\varepsilon_{int} + \varepsilon_{ext}\right)$\cite{keldysh_coulomb_1979}. This background screening function interpolates as a function of the effective material thickness $h$ between the external dielectric screening $\varepsilon_{ext}$, such as arising from surrounding material, in the long-wavelength limit and the internal interband screening $\varepsilon_{int}$ in the short-wavelength limit. For $h \rightarrow 0$ we find $\varepsilon(q) \rightarrow \varepsilon_{ext}$, such that internal background screening becomes negligible. 
This model has been shown to adequately describe all non-metallic screening channels in ab initio constrained Random Phase Approximation (cRPA) calculations for various layered materials~\cite{rosner_wannier_2015,steinhoff_influence_2014,schonhoff_interplay_2016,steinke_coulomb-engineered_2020,soriano_environmental_2021,stepanov_coexisting_2022} and was also used for the description of force constants in layered materials~\cite{royo_exact_2021}. $v_C(q)$ may thus be tuned in experiments by varying the substrate on which the sample is mounted and the materials that encapsulate it. For simplicity we further consider here dispersionless longitudinal (LO) and transverse (TO) optical phonons with the same phonon frequency $\omega_e = 0.3$\,eV, and momentum-independent electron-phonon coupling $g^2 = 0.3$\,eV$^2$, such that the bare phonon-mediated electron-electron interactions read $v_{ph}^{LO/TO}(\inu_n) = g^2 2 \omega_e / [(\inu_n)^2 - \omega_e^2]$. In the Supplemental Material we further show that non-local electron-phonon coupling models do not affect our conclusions.
We finally use the random phase approximation (RPA) to compute the mutual screening of electrons and longitudinal phonon modes (the transverse mode is unaffected by the Coulomb interaction) as arising from the metallic states. The longitudinal interaction then reads
\begin{equation}
    \label{eq:mutualScreening}
    I_L(q, \inu_n) = \frac{v_C(q) + v_{ph}^{LO}(\inu_n)}{1 - [v_C(q) + v_{ph}^{LO}(\inu_n) ] \Pi_0(q, \inu_n)},
\end{equation}
with $\Pi_0(q, \inu_n)$ the RPA polarization. The full interaction $I_0(q, \inu_n)= I_L(q, \inu_n) + v_{ph}^{TO}(\inu_n)$ is the sum of the longitudinal and transverse interactions. The resulting interaction hosts frequency and momentum dependent screened electron-phonon~\cite{ponce_long-range_2022,berges_phonon_2022,girotto_dynamical_2023} and electron-electron interactions as well as their hybridization.
We evaluate the electron dispersion within a nearest-neighbour square-lattice tight binding model with a hopping of $t = 1.5$\,eV ($m^* \approx 0.1$\,$m_e$) and a chemical potential yielding approximately quarter filling. $n$ is such that the electron gas parameter $r_s = m^* e^2 / (\varepsilon_{ext} \sqrt{\pi n}) < 1$ for all values of $\varepsilon$ considered, justifying the RPA~\cite{takada_electron_1991,mahan_many-particle_2000}.
For later reference we define the statically screened 
\begin{equation}
    \label{eq:staticMutualScreening}
    I^{stat}_L(q, \inu_n) = \frac{v_C(q) + v_{ph}^{LO}(\inu_n)}{1 - [v_C(q) + v_{ph}^{LO}(\inu_n) ] \Pi_0(q, \inu_n=0)}
\end{equation}
and the non-mutually screened
\begin{align}
    \label{eq:nonMutualScreening}
    I^{C+ph}_L(q, \inu_n) &= \frac{v_C(q)}{1 - v_C(q) \Pi_0(q, \inu_n)}\\
    &+
    \frac{v_{ph}^{LO}(\inu_n)}{1 - v_{ph}^{LO}(\inu_n) \Pi_0(q, \inu_n)} \notag
\end{align}
longitudinal interactions.
Comparison of results obtained with these model interactions to results obtained with the full interaction allows us to study the effects of plasmons, which are absent in $I^{stat}_L(q, \inu_n)$, and the mutual electron-phonon screening, which is neglected in $I^{C+ph}_L(q, \inu_n)$.

\begin{figure}
    \includegraphics[width=0.99\columnwidth]{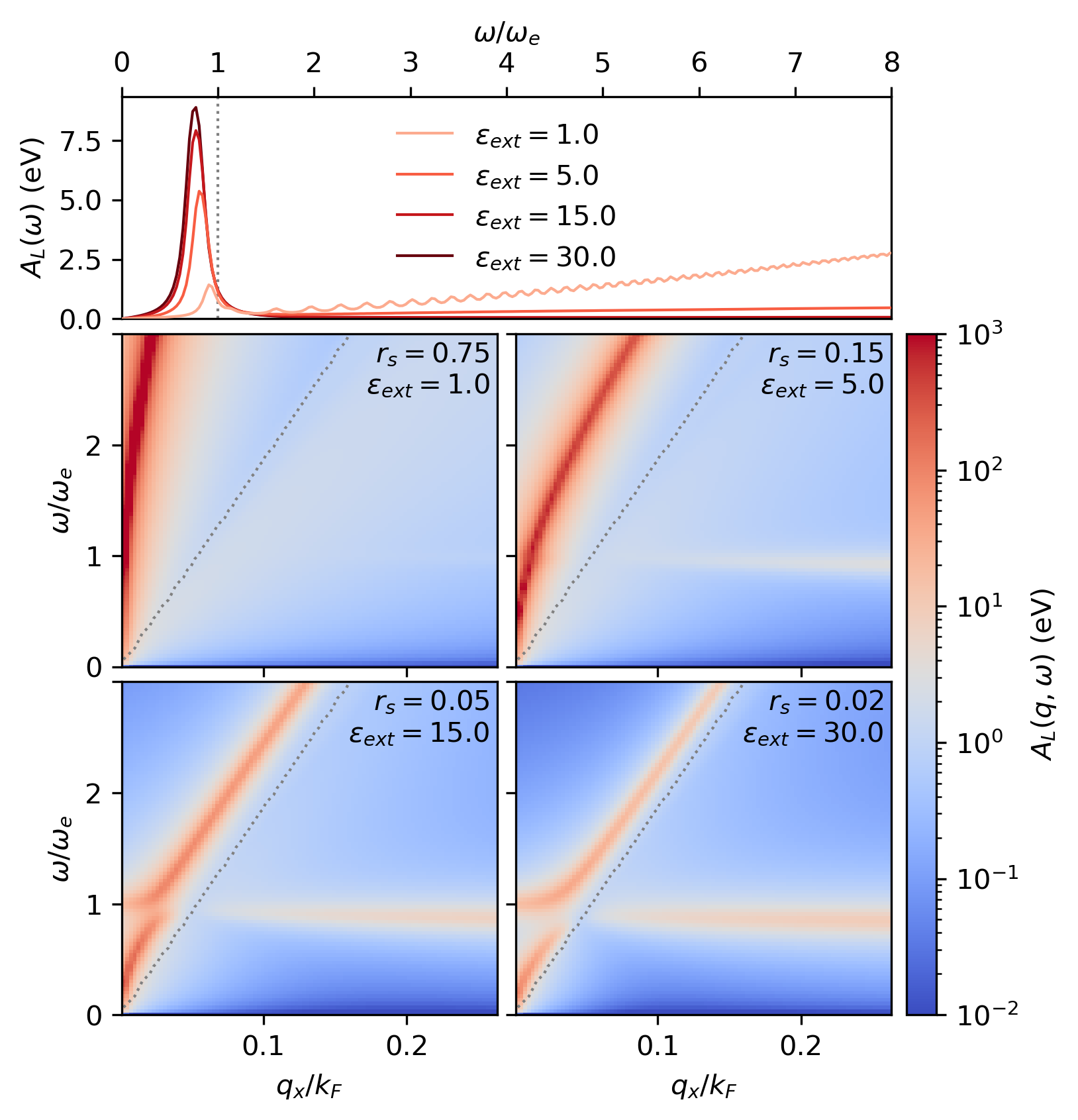}
    \caption{\label{fig:EELS} Momentum-integrated longitudinal interaction spectral function $A_L(\omega) = \sum_{q\neq0} A_L(q,\omega)$ (top panel) and momentum-resolved longitudinal spectral function $A_L(q,\omega)$ (bottom panels) of the mutually screened longitudinal interaction $I_L(q, \omega)$, calculated using Eq.~(\ref{eq:mutualScreening}) on the real frequency axis for various $\varepsilon_{ext}$ in the $h\rightarrow0$ limit}. The dotted gray line in the bottom panels traces the upper frequency boundary of the electron-hole continuum.
\end{figure}

The real-frequency spectral functions of the longitudinal interaction $A_L(q,\omega) = -\text{Im}\left( I_L(q, \omega) \right)$, together with their local spectra $A_L(\omega) = \sum_{q\neq0} A_L(q,\omega)$, are shown in Fig.~\ref{fig:EELS} for a variety of $\varepsilon_{ext}$ (in the $h\rightarrow0$ limit). For clarity we present only the longitudinal part, as the transverse part will simply add a dispersionless peak at $\omega = \omega_e$. For $\varepsilon = 1$ we find the $\sqrt{q}$-like 2D plasmon dispersion. In this regime the bare Coulomb interaction strongly screens the electron-phonon interaction, causing the latter to be negligible. As a result, $A_L(\omega)$ is governed by the plasmon spectrum. As $\varepsilon$ is increased, the electron-phonon interaction is screened less, and the phonon dispersion starts to appear at large $q$. The effective phonon frequency is reduced in comparison to the bare $\omega_e$ as a result of screening. This is also visible in $A_L(\omega)$, showing phonon peaks always below $\omega_e$, which gain intensity as $\varepsilon$ decreases.
In addition we find that the phonons and plasmons start to hybridize, creating a gap between the two branches around $q \approx 0.05 k_F$, with $k_F$ the Fermi wave vector. Due to this non-trivial momentum structure of the interaction a consistent theory must retain both the full frequency and momentum structure of the normal and anomalous self-energies to describe superconductivity in 2D materials.

\textbf{Superconductivity.} We treat both the normal and anomalous self-energies in a one-loop approximation. The linearized equation for the  anomalous self-energy $\phi(k, \iw_n)$ is an eigenvalue equation that may be written
\begin{align}
    \label{eq:gapEqn}
    \lambda(T)\phi(k, \iw_n) &= 
    -\frac{1}{\beta} 
         \sum_{k' \neq k, m}
           I_0(k-k', \iw_n - \iw_m) \\
           &\times G_e(k', \iw_m) \phi(k', \iw_m) G_e(-k', -\iw_m), \notag
\end{align}
where $\beta$ is the inverse temperature. The dressed normal-state Green's function $G_e(k, \iw_n)$ is obtained from $G_e(k, \iw_n)^{-1} = G_0(k, \iw_n)^{-1} - \Sigma(k, \iw_n)$, with $G_0(k, \iw_n)$ the bare Green's function and $\Sigma(k, \iw_n)$ the normal state self-energy given by
\begin{equation}
    \Sigma(k, \iw_n) = 
        -\frac{1}{\beta}
         \sum_{k' \neq k, m}
           I_0(k-k', \iw_n - \iw_m)
           G_0(k', \iw_m).
\end{equation}
In Eq.~(\ref{eq:gapEqn}) we include the eigenvalue $\lambda(T)$ on the left-hand side such that the transition temperature $T_c$ is defined by $\lambda(T_c) = 1$. The term $k=k'$ is excluded to account for the counteracting positive charge background, which is present in any realistic system. The gap equation was solved using an iterative solver implemented in the TRIQS~\cite{parcollet_triqs_2015} and TPRF~\cite{wentzell_triqstprf_2022} packages using $180\times180$ $k$ and $q$ grids and a Matsubara cut-off of $\omega_c = 30\,$eV.

\textbf{Results: Transition temperature.} Resolving low $T_c$ is challenging due to the number of required Matsubara frequencies. Instead we analyze the leading eigenvalue $\lambda$ at a fixed temperature $T = 98$\,K,  which is the critical temperature for $\varepsilon_{ext} \rightarrow \infty$, i.e., without contributions from the Coulomb interaction. This value is for reference only. The critical temperatures for the following calculations taking the full Coulomb interaction effects into account range between $6$K and $60$K. The dependence of $\lambda$ is a proxy for the dependence of $T_c$ on parameters.

\begin{figure*}
    \includegraphics[width=0.99\textwidth]{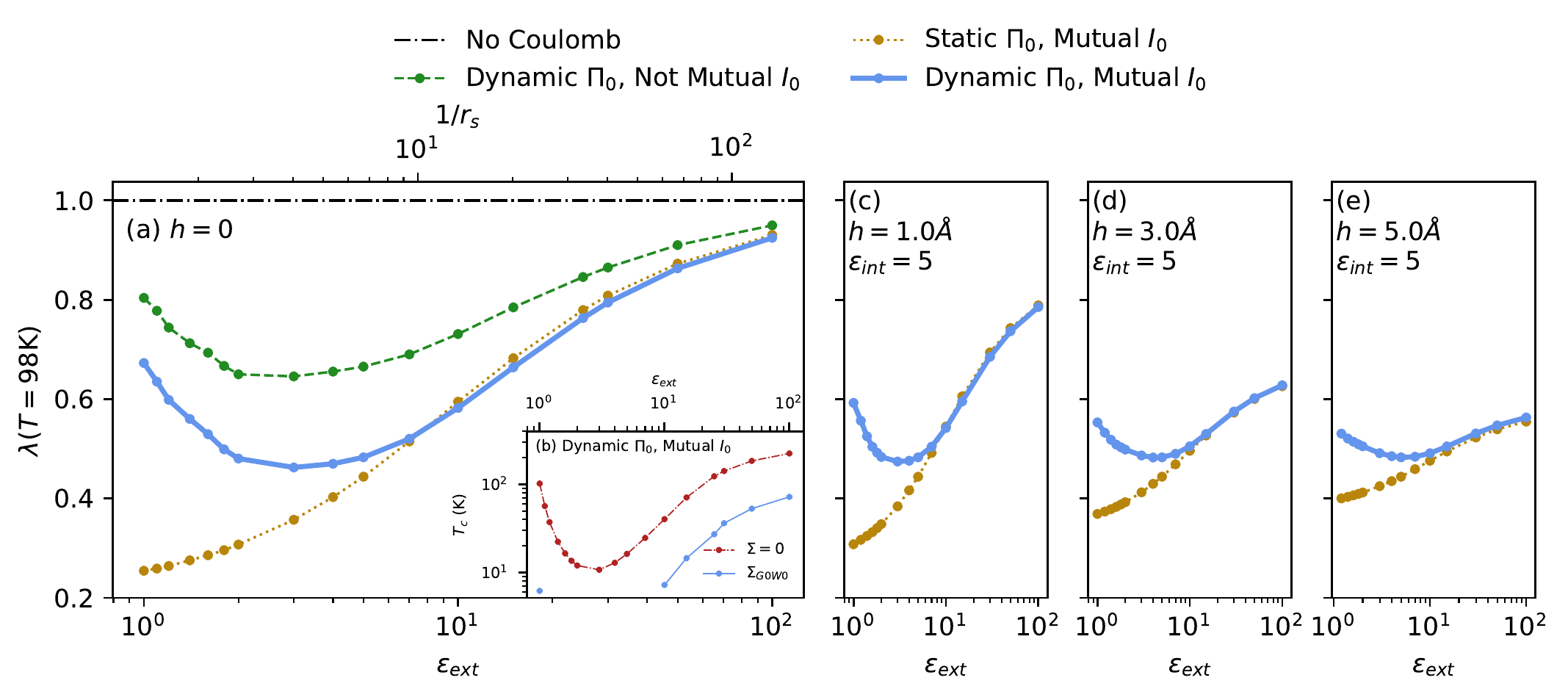}
    \caption{\label{fig:leadingEv}(a) Leading eigenvalue $\lambda$ of Eq.~(\ref{eq:gapEqn}) at temperature $T = T_c^{\varepsilon \rightarrow \infty} = 98$\,K as a function of the external background screening parameter $\varepsilon_{ext}$ at $h=0$ (lower axis) and the inverse gas parameter $1/r_s = \left(m^* e^2 / (\varepsilon_{ext} \sqrt{\pi n})\right)^{-1}$ (upper axis), for a variety of different models for the total interaction. 
    (Black-dashed) Neglecting the Coulomb interaction, i.e., $I_0(q, \inu_n)$ where $v_C(q) = 0$.
    (Green-dashed) Neglecting mutual screening $I^{C+ph}_0(q, \inu_n)$.
    (Yellow-dotted) Statically mutually screened $I^{stat}_0(q, \inu_n)$.
    (Blue-solid) Fully dynamically and mutually screened $I_0(q, \inu_n)$. 
    (b) The critical temperatures for the full dynamic and mutually screened model obtained by solving the linearized gap equation including (blue) and excluding (red) normal state renormalization. (c)-(e) The leading eigenvalue $\lambda$ at $T = 98$\,K as a function of external screening $\varepsilon_{ext}$, for a variety of effective material thicknesses $h$ at $\varepsilon_{int} = 5$.}
\end{figure*}

The solid-blue curve in Fig.~\ref{fig:leadingEv}(a) shows $\lambda$ of as a function of the strength of the Coulomb interaction controlled by the background dielectric constant $\varepsilon_{ext}$ in the $h\rightarrow0$ limit. As $\varepsilon_{ext}\rightarrow\infty$ the Coulomb interaction drops out and the eigenvalue tends to the value $\lambda=1$ found for $T = 98$\,K in the phonon-only model. 
Decreasing $\varepsilon_{ext}$ from $\infty$ initially decreases $T_c$ consistent with the conventional expectation that the Coulomb repulsion formally counteracts the phonon-mediated attraction. 
However, a minimum in $\lambda$ at $\varepsilon_{ext} \approx 3$ is evident, and for smaller $\varepsilon_{ext}$ the leading eigenvalue again increases, signalling a different behavior in the Coulomb-dominated small $\varepsilon_{ext}$ regime.

In Fig.~\ref{fig:leadingEv}(a) we compare the full $\lambda$ to that found in the phonon-only limit (black-dashed line), from the statically screened interaction $I^{stat}_0(q, \inu_n)$ (orange-dotted line), and the dynamical interaction without mutual screening between the electron-electron and electron-phonon interactions $I^{C+ph}_0(q, \inu_n)$ (green-dashed line). We first analyse the weak Coulomb regime $\varepsilon_{ext} \gtrsim 7$. Here, we see that dynamic screening has a negligible effect on superconductivity (orange and blue lines coincide) but the static Coulomb interaction is important (difference from black line). In this regime the plasmon is strongly Landau-damped, rendering the electron-phonon interaction the dominant pairing interaction. The superconducting state can qualitatively be understood within the Tolmachev-Morel-Anderson pseudopotential $\mu^*$ approximation~\cite{tolmachev_logarithmic_1961,morel_calculation_1962}, such that the transition temperature scales exponentially as $e^{-1/\lambda^\star}$, with the effective pairing strength $1/\lambda^\star=(1 + \lambda_{ph}) / (\lambda_{ph} - \mu^*)$~\cite{mcmillan_transition_1968,allen_transition_1975,dynes_mcmillans_1972}. As $\varepsilon_{ext}$ is decreased the Coulomb repulsion parameterized by $\mu^*$ increases in strength, leading to a lower $T_c$. 
Furthermore, we find in this regime that $\lambda$ is overestimated when the cross-coupling of the longitudinal phonon and the plasmon is neglected (difference between green and orange/blue lines). This effect can be understood as an effective decrease of $\lambda_{ph}$, due to the static Coulomb interaction screening the electron-phonon interaction~\cite{simonato_revised_2022,akashi_revisiting_2022,pellegrini_eliashberg_2022}.

In the $\varepsilon_{ext} < 3$ regime we see that the static screening approximation (orange line) predicts a strongly suppressed $T_c$ compared to the full model (blue line). This is due to the static Coulomb interaction which strongly screens the electron-phonon interaction, rendering conventional phonon pairing negligible. From this we can conclude that in this regime the dominant pairing channel electron-plasmon interactions. This is in line with the spectral functions shown in Fig.~\ref{fig:EELS}, where the phonon mode is negligible compared to the plasmon mode at $\varepsilon_{ext} = 1$. Mutual screening between the different bosons again decreases $\lambda$ (difference between green and blue curves), which we still understand as resulting from the strong screening of the electron-phonon interactions in this limit. 

The minimum around $\varepsilon_{ext} \approx 3$ arises from the crossover between the two different pairing regimes, which vanish in opposite limits. Here, the electron-plasmon coupling is too weak and the static Coulomb repulsion too strong to induce any strong pairing.

\begin{figure*}[t]
    \includegraphics[width=0.99\textwidth]{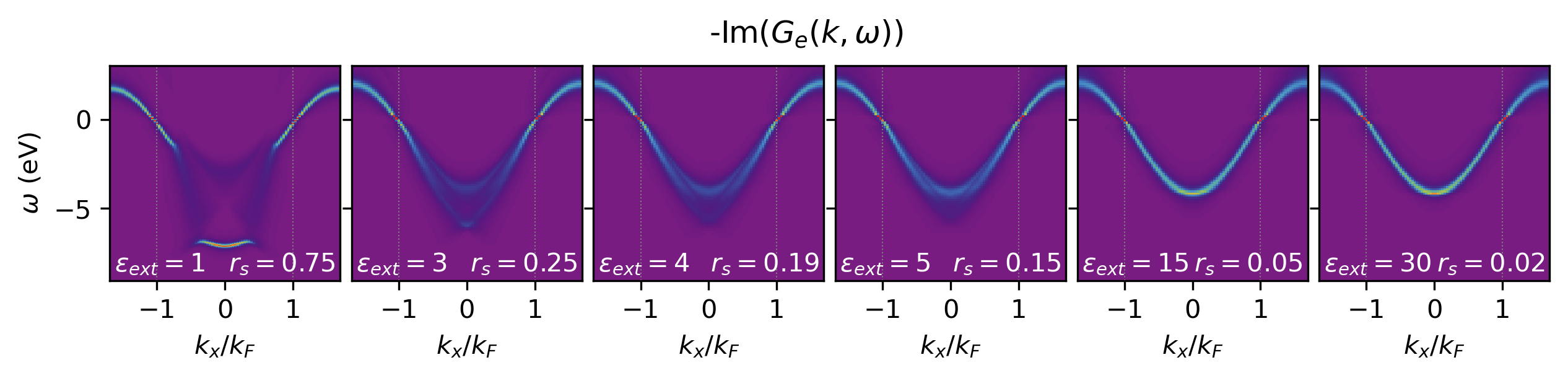}
    \caption{\label{fig:SpecFunc} Spectral function of the dressed normal-state Green's function $-\text{Im}\left( G_e(k,\iw_m) \right)$ calculated on the real frequency axis for various $\varepsilon_{ext}$.}
\end{figure*}

In the inset of Fig.~\ref{fig:leadingEv}(a) we compare the superconducting transition temperature $T_c$ of the full model to $T_c$ calculated from Eq.~(\ref{eq:gapEqn}) without the normal-state self energy. For the full model, the computational complexities restrict us to $T_c$ only over part of the $\varepsilon_{ext}$ range, but in the model without normal state self-energy the entire curve can be traced out. These curves confirm that the $\varepsilon_{ext}$ dependence of the leading eigenvalue is a good proxy for the $\varepsilon_{ext}$ dependence of the transition temperature. We further see, in agreement with results previously obtained in the 3D plasmon-only model~\cite{davydov_ab_2020,wang_origin_2022}, that normal-state self-energy effects drastically reduce $T_c$, which is in 2D, however,  strongly tuned by the screening $\varepsilon$.

In Figs.~\ref{fig:leadingEv}(c)-(e) we further show the leading eigenvalue $\lambda$ as a function of external screening $\varepsilon_{ext}$, at fixed internal screening $\varepsilon_{int}$ for various finite material thicknesses $h$. We again find the same qualitative behaviour with distinct plasmon and phonon mediated regimes at small and large $\varepsilon_{ext}$, respectively. As the material thickness $h$ increases the superconducting state becomes less susceptible to the external screening, as the internal screening $\varepsilon_{int}$ becomes more and more dominant (see SM for more details).

\textbf{Normal State.} In Fig.~\ref{fig:SpecFunc} we show the spectral function of $G_e(k, \omega)$ computed directly on the real-frequency axis and for $h\rightarrow0$. For $\varepsilon_{ext}=30$ and $\varepsilon_{ext}=15$ we find the conventional electron-phonon coupling induced mass enhancement, accompanied by an increase of the spectral weight at the bottom of the quasi-particle band. Both of these features are characteristic for phononic normal-state renormalizations.~\cite{giustino_electron-phonon_2017} This is in line with the decrease of $T_c$ by a factor of about $3$ for $\varepsilon \geq 7$, as shown in the inset of Fig.~\ref{fig:leadingEv}, which is well approximated by the ratio of $T_c$ with and without normal-state contributions $T_c(\Sigma)/T_c(\Sigma=0) \approx e^{\frac{1}{\lambda}-\frac{1+\lambda}{\lambda}}=e^{-1}\approx 1/2.72$.

As $\varepsilon_{ext}$ is decreased below $5$, spectral weight of occupied states around $k=0$ is shifted away from the main quasi-particle band into a plasmonic side band at lower energies. Eventually, at $\varepsilon_{ext} = 1$, due to the strong static Coulomb interaction, most of the spectral weight has shifted into the shake-off band, rendering the latter strongly coherent and the initial quasi-particle band incoherent. These results are reminiscent of plasmonic polarons, which were observed in various 2D materials using angle-resolved photoemission spectroscopy measurements.~\cite{bostwick_observation_2010,caruso_two-dimensional_2021} At $\varepsilon_{ext} = 1$, $T_c$ is reduced by a factor of $18$ by the normal-state self-energy, as visible in the inset of Fig.~\ref{fig:leadingEv}. This significant $T_c$ reduction is caused by the strong spectral weight transfer, which can be understood within the quasi-particle approximation $G_e(k, \omega) \approx Z_k G_0(k,\omega)$, where $Z_k^{-1} = 1 - \partial_\omega \Sigma(k, \omega) \vert_{\omega = \xi_k}$ is a measure for the amount of spectral weight transfer and $\xi_k$ is the bare electron dispersion. This approximation is justified in the low $\varepsilon_{ext}$ regime, as here the dynamic and non-local Coulomb interaction does not significantly change the effective mass.~\cite{akashi_revisiting_2022} Within this approximation it is clear that the anomalous self-energy of Eq.~(\ref{eq:gapEqn}) is scaled by a factor $Z_k^2 < 1$, yielding a strongly reduced $T_c$.~\cite{davydov_ab_2020,wang_origin_2022}

\begin{figure}
    \includegraphics[width=0.99\columnwidth]{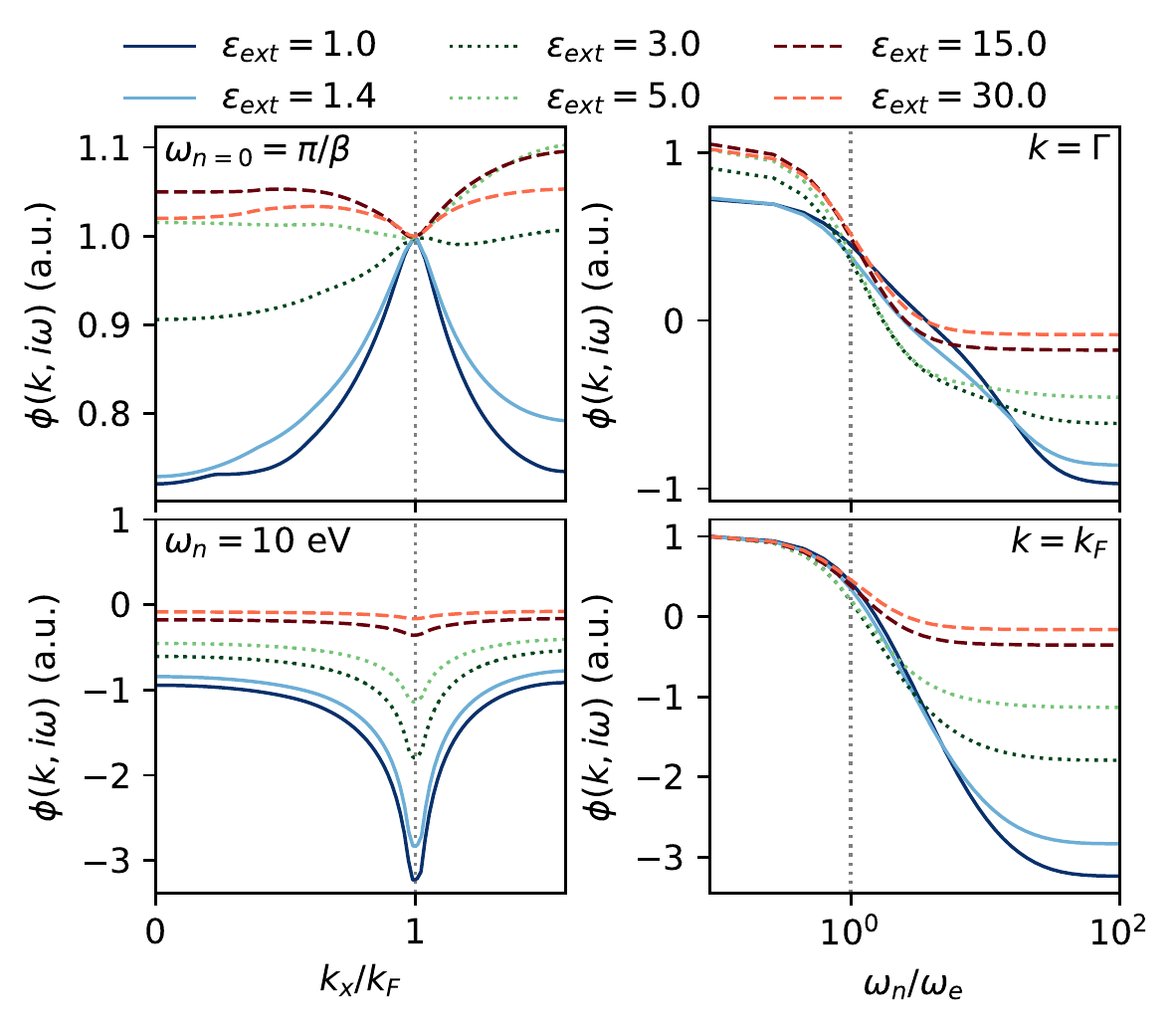}
    \caption{\label{fig:Delta} Eigenvector $\phi(k, \iw_n)$ corresponding to the leading eigenvalue of Eq.~(\ref{eq:gapEqn}) at temperature $T = T_c^{\varepsilon \rightarrow \infty} = 98$\,K, as a function of $k_x$ (left panels) and as a function of Matsubara frequency $\omega_n$ (right panels), for $\varepsilon_{ext}$ in the plasmonic (blue solid), phononic (red dashed) and intermediate (green dotted) regimes. In all cases $\phi(k, \iw_n)$ is scaled such that $\phi(k=k_F, \iw_{n=0}) = 1$.}
\end{figure}

\textbf{Anomalous Self-Energy.} The distinct phonon and plasmon mediated regimes are also clearly visible in the eigenvectors $\phi(k, \iw_n)$, shown in Fig.~\ref{fig:Delta}. 
In the phonon-mediated regime ($\varepsilon_{ext} > 3$), $\phi(k, \iw_n)$ shows the conventional characteristics of electron-phonon mediated superconductivity affected by static Coulomb repulsion: we find a strong peak around $\omega_{n=0}$ with a width on the order of the phonon frequency $\omega_e$ accompanied by a negative high-frequency tail. In momentum space $\phi(k, \iw_n)$ is only weakly structured as a result of the momentum-independent bare electron-phonon interaction $v_{ph}(\inu)$ in our model. Approximating the Coulomb interaction with a constant Tolmachev-Morel-Anderson pseudopotential $\mu^*$ might thus be reasonable for phonon mediated superconductivity in this regime.

In the plasmon-mediated regime ($\varepsilon_{ext} < 3$), $\phi(k, \iw_n)$ also shows a peak around $\omega_{n=0}$, but with an enhanced width, which increases with decreasing $\varepsilon$. The high-frequency negative tail is furthermore strongly enhanced. In momentum space, $\phi(k, \iw_n)$ has a strong momentum dependence around $k_F$, which is qualitatively changing with Matsubara frequency. While $\phi(k, \iw_n)$ shows a maximum at $\iw_{n=0}$, it turns into a minimum at large $\iw_n$, which we identify as a clear signature of plasmonic superconductivity in 2D. In the plasmon-dominated limit at $\varepsilon_{ext} = 1$, we understand this behaviour as a result of the interplay between the static bare and dynamic RPA screened Coulomb interaction $W(q,\omega) = v_C(q)/[1 - v_C(q) \Pi_0(q,\omega)] = v_C(q) + \Delta W(q,\omega)$, which, respectively, yield a repulsive static $\phi_C(q)$ and attractive dynamic $\phi_{\Delta W}(q,\omega)$ in analogy to the conventional electron-phonon paring mechanism under the influence of static Coulomb repulsion. Here, however, the dyanmic $\phi_{\Delta W}(q,\omega)$ is controlled by the rather large plasmon energies, which does not allow for a separation of energies anymore. As a result there is no logarithmic renormalization of the repulsive term, which is furthermore driven by the bare Coulomb interaction $v_C(q)$ rather than by the statically screened one $W(q,\omega=0)$. This explains the strong negative tail in frequency space. Since the attractive $\phi_{\Delta W}(q,\omega)$ is driven by the plasmon frequencies the pairing frequency is enhanced. Finally, as the attraction is induced by the electron-plasmon coupling, which strongly favours pairing at small momentum transfer $q$, we find a pronounced momentum structure.
We also note that in the plasmon-mediated regime the system is very sensitive to the precise value of $\varepsilon_{ext}$, which is inline with the Coulomb-based interpretation. The negative high-frequency tail of $\phi(k, \iw_n)$ is, e.g., more than halved from $\varepsilon_{ext}=1$ to $\varepsilon_{ext}=5$ and the leading eigenvalue drops from $\lambda \approx 0.67$ to $\lambda \approx 0.48$.

In the intermediate regime around $\varepsilon_{ext}\approx3$ we find a pronounced negative tail in the frequency domain, which cannot be compensated by the weak momentum structure. This again shows the interplay between the weak electron-plasmon coupling and strong static Coulomb repulsion, which qualitatively explains the suppressed superconducting state here.

\textbf{Conclusions.} 
We have shown that superconductivity in layered materials, when consistently influenced by electron-phonon, electron-plasmon, and phonon-plasmon interactions, can develop three distinct regimes, which can be tuned by the overall strength of Coulomb interaction. In all cases considered here $r_s < 1$ holds, which is different to the seminal results by Takada obtained for $r_s > 1$.~\cite{takada_plasmon_1978,takada_plasmon_1992} For strong Coulomb interactions (for small internal and external screening) we find a plasmon mediated regime, for (strongly) reduced Coulomb interactions we find that phonon mediation becomes prominent, while for intermediate Coulomb interaction the static repulsion is too strong and the plasmonic pairing too weak such that superconductivity is overall suppressed. These regimes have clear individual footprints in the gap functions and also show distinctively different impacts of the normal-self energy renormalizations. We can expect to find these phases for materials with a small internal background screening $\varepsilon_{int}$ and can expect the external, e.g., substrate-screening, tunability to be most prominent if the effective height $h$ of the material is rather small.

Our results furthermore show that two-dimensional superconductivity mediated by plasmons is possible within the RPA when the background screening is weak, such that the plasmon dispersion is well separated from the electron-hole continuum. In this regime the electron-phonon pairing is mostly suppressed due screening.  While above we focused on the sensitive role of the screening $\varepsilon_{ext}$, we show in the supplement that the plasmonic regime also survives as a function of doping, with distinctively different characteristics from the phonon-mediated regime.
Our results are qualitatively similar to the findings by Wang \textit{et al.} in 3D~\cite{wang_origin_2022}. However, here we consistently account for the mutual screening between electrons and phonons, find a minimum in $T_c$ around $r_s \approx 0.25$, and have an overall more pronounced structure in $T_c(r_s)$, which we find to be strongly influenced by the environmental screening. We attribute these differences between 3D and 2D to the qualitatively different characteristics of plasmonic excitations in 2D. 

The pronounced sensitivity to the background screening which generates the plasmon- and phonon-mediated and intermediate regimes, is important for the descriptions nd measurements of layered superconducting heterostructures. Most experimental data on layered superconductors are obtained from samples mounted on some substrates or encapsulated by other layered materials, which creates a non-trivial screening environment as described by our background-dielectric function. Our results also show that this background screening channel can be utilized to precisely tune the superconducting state via modifications of the environment. 
In the limit of strong substrate screening phonons are responsible for Cooper pairing. In this regime our results show that the Coulomb pseudo-potential $\mu^*$ can be a valid approximation. However, mutual screening should be taken into account, as neglecting it will overestimate the critical temperature, and the residual static but non-local Coulomb interaction must be handled carefully~\cite{simonato_revised_2022,akashi_revisiting_2022,pellegrini_eliashberg_2022}.

The intermediate regime around $\varepsilon_{ext} = 3$, in which both electron-plasmon and electron-phonon interactions act simultaneously, is likely especially relevant. Here the $\mu^*$ approximation can cause an underestimation of the critical temperature and would hide plasmonic footprints on the gap function. As discussed in the supplement, a signature of this situation could be that fits of $\mu^*$ to experimental results yield unexpected trends as a function of $\varepsilon$ or the doping of a system.

\textbf{Acknowledgement.} Y.I.V. thanks the Flatiron Institute's Predoc program for hospitality and support. Y.I.V,  M.R., and M.I.K. acknowledge support from the Dutch Research Council (NWO) via the “TOPCORE” consortium. M.R., and M.I.K. acknowledge the research program “Materials for the Quantum Age” (QuMat) for financial support. This program (registration number 024.005.006) is part of the Gravitation program financed by the Dutch Ministry of Education, Culture and Science (OCW). 
The work of M.I.K. was further supported by the European Union’s Horizon 2020 research and innovation program under European Research Council Synergy Grant 854843 “FASTCORR”. A.J.M. was supported in part by  A. J. M. acknowledge support from the National Science Foundation (NSF) Materials Research Science and En- gineering Centers (MRSEC) program through Columbia University in the Center for Precision Assembly of Super- stratic and Superatomic Solids under Grant No. DMR- 1420634 and Grant No. DMR-2011738. The Flatiron Institute is a division of the Simons Foundation. 

\newpage
\onecolumngrid
\section{Supplemental Material}

\subsection{Doping Dependence}

\begin{figure}[h]
    \includegraphics[width=0.99\textwidth]{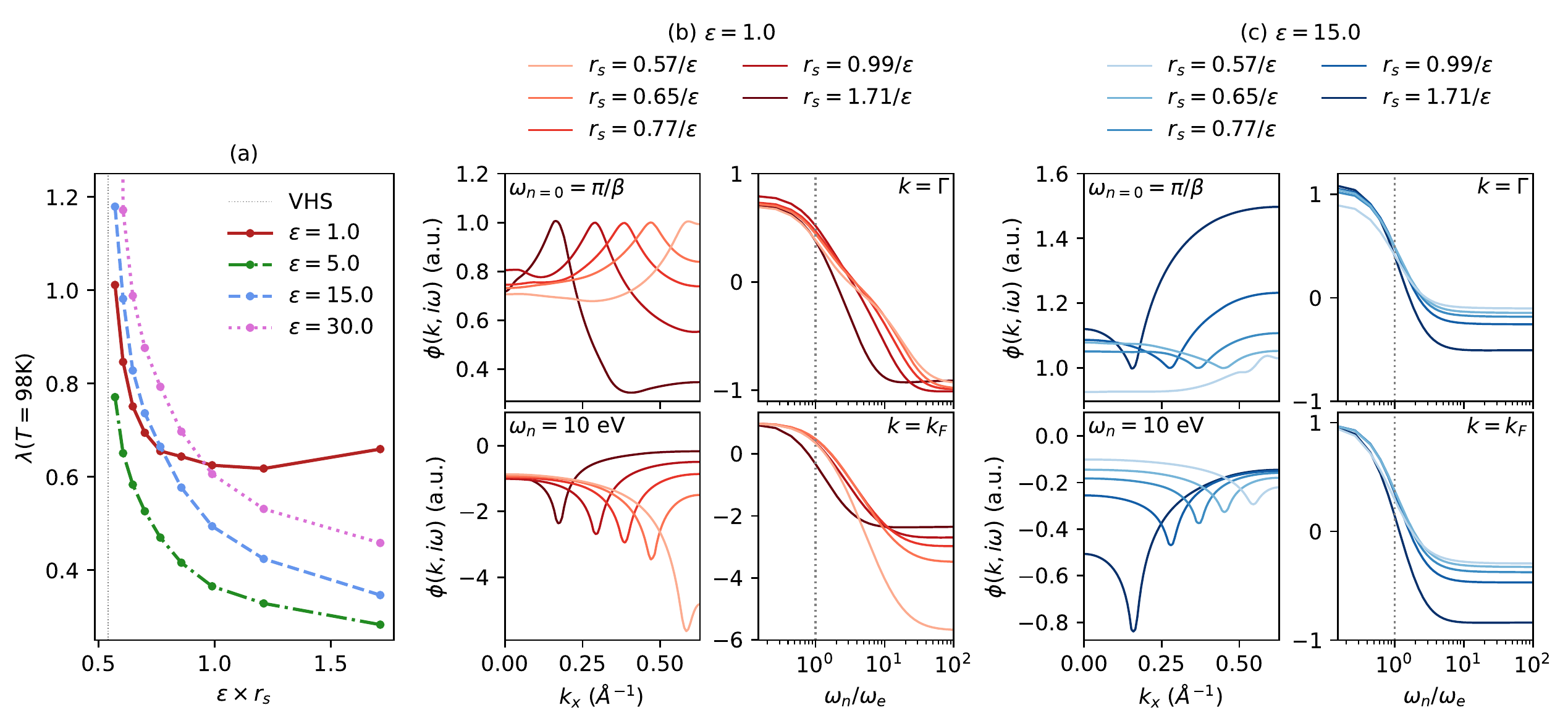}
    \caption{\label{fig:varyingDoping} (a) The leading eigenvalue $\lambda$ of Eq.~(4) at temperature $T = 98$\,K for the fully mutual and dynamic model $I_0(q, \inu_n)$, as a function of $\varepsilon \times r_s$, while varying the electron occupation, for various $\varepsilon$. The dotted gray line indicates the location of the Van Hove singularity.
    The other panels show the corresponding eigenvectors $\phi(k, \iw_n)$ at $\varepsilon = 1$ (b) and $\varepsilon = 15$ (c) at temperature $T = 98$\,K, as a function of $k_x$ and as a function of Matsubara frequency $\omega_n$, for various electron occupations. In all cases $\phi(k, \iw_n)$ was scaled such that $\phi(k=k_F, \iw_{n=0}) = 1$.}
\end{figure}

To verify the robustness of our results with respect to the doping level, we performed additional calculations while varying the electron occupation $n$ and present the results in Fig.~\ref{fig:varyingDoping} as a function of the electron gas parameter $r_s = m^* e^2 / (\varepsilon \sqrt{\pi n})$. In Fig.~\ref{fig:varyingDoping}(a) we show the leading eigenvalue $\lambda$ while varying the electron occupation for the full dynamically mutually screened model $I_0(q, \inu_n)$. 
For the $\varepsilon$ considered, $\lambda$ is smallest at $\varepsilon = 5$ for all electron occupations, indicating that the trends with $\varepsilon$ discussed in the main text are robust under varying electron occupation.
For all $\varepsilon$ at small $r_s$ we find signatures in line with electron-phonon mediated superconductivity. As the doping is increased towards the van Hove singularity (around $r_s \approx 0.5 / \varepsilon$), the density of states at the Fermi level increases, and therefore the electron-phonon mediated $T_c$ increases as well due to the enhanced $\lambda^*$.
At large $r_s$ and $\varepsilon = 1$ we find an increase of $\lambda$, whereas for the other $\varepsilon$ $\lambda$ decreases, which we again identify as the plasmon-mediated regime. 

In Fig.~\ref{fig:varyingDoping} (b) and (c) we analyze the eigenvectors $\phi(k, \iw_n)$ for $\varepsilon = 1$ (plasmon regime) and $\varepsilon = 15$ (phonon regime), respectively. The trends of the eigenvector with occupation are rather different in the two regimes. At $\varepsilon = 15$ the results can be understood using the conventional electron-phonon and static Coulomb interactions. As $r_s$ is increased away from the Van Hove singularity (doping is reduced), the screening is reduced and thus the Coulomb interaction is enhanced. This is reflected in $\phi(k, \iw_n)$ by the enhanced high-frequency tail and by the more pronounced momentum structure at large $r_s$.
At $\varepsilon = 1$ we find that the negative high-frequency tail at $k_F$ is reduced as $r_s$ increases, which is opposite to what one would expect from the Tolmachev-Morel-Anderson Coulomb pseudopotential approximation $\mu^*$~\cite{tolmachev_logarithmic_1961,morel_calculation_1962}, showing again the breakdown of this approximation in the weakly screened limit. We also find that, unlike at $\varepsilon = 15$, the strong momentum peak of $\phi(k, \iw_n)$ at $\varepsilon = 1$ and $\omega_n=10$\,eV is reduced as $r_s$ increases. Simultaneously the momentum structure at $\iw_{n=0}$ is enhanced, such that we see a shift of the momentum structure from $\iw_n=10$\,eV to $\iw_{n=0}$ as $r_s$ increases at $\varepsilon = 1$.
A final feature that highlights the distinct phonon and plasmon regimes is the width of the peak of $\phi(k, \iw_n)$ along $\omega_n$. At $\varepsilon = 15$ the width is close to the bare phonon frequency $\omega_e$ for all $r_s$, showing that the dominant pairing boson is the phonon at all doping levels. On the other hand, at $\varepsilon = 1$ the width of $\phi(k, \iw_n)$ at $k=\Gamma$ decreases as $r_s$ increases. This indicates that the energy of the pairing boson decreases with decreased doping, as expected for the two-dimensional plasmon dispersion.

\subsection{Non-Local Electron-Phonon Interaction}

\begin{figure}
    \centering
    \includegraphics[width=0.6\textwidth]{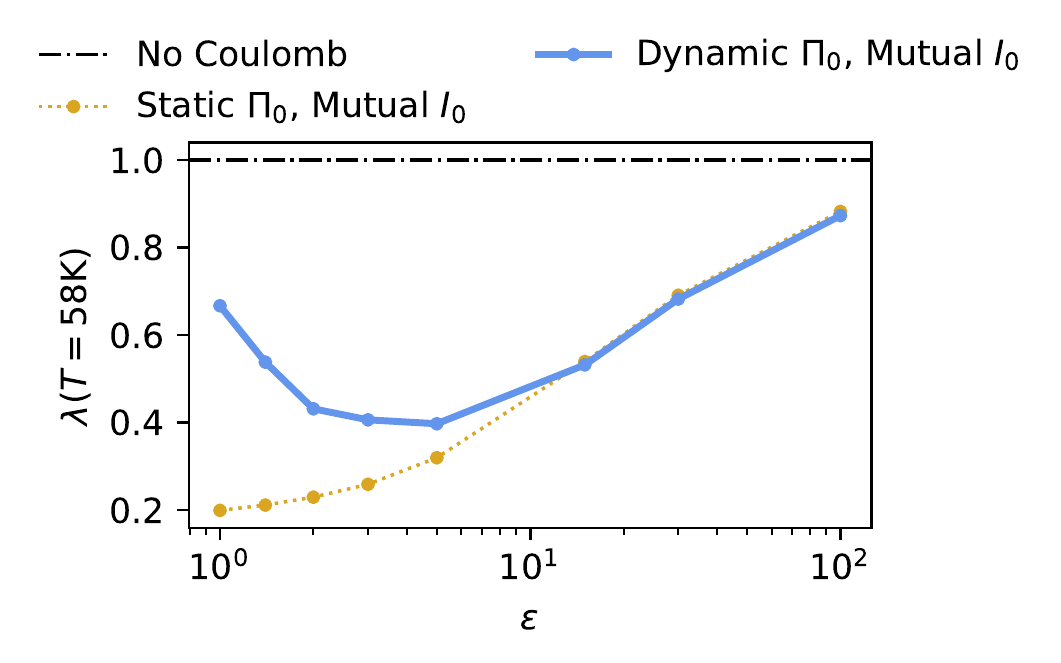}
    \caption{\label{fig:nonlocalPhonon}
    Leading eigenvalue $\lambda$ at temperature $T = T_c^{\varepsilon \rightarrow \infty} = 58$\,K as a function of the local background screening parameter $\varepsilon$, using a non-local Fr\"ohlich electron-phonon interaction $g(q)^2$ with parameters $h = 2$\,\AA, $g_F^2 = 0.7$\,eV$^2$ and $\omega_e = 0.3$\,eV, for a variety of different models for the total interaction.
    (Black-dashed) Neglecting the Coulomb interaction, i.e., $I_0(q, \inu_n)$ where $v_C(q) = 0$.
    (Yellow-dotted) Statically mutually screened $I^{stat}_0(q, \inu_n)$.
    (Blue-solid) Fully dynamically and mutually screened $I_0(q, \inu_n)$. 
    }
\end{figure}

In order to understand how non-locality in the bare electron-phonon coupling will affect the results of the main text, we consider the following Fr\"olich electron-phonon coupling appropriate for 2D materials\cite{kaasbjerg_phonon-limited_2012}
\begin{equation}
    g(q)^2 = g_F^2 \text{erfc}\left(\frac{1}{2}hq\right)^2,
\end{equation}
where $\text{erfc(x)}$ is the complementary error function and $h$ tunes the non-locality induced by the effective thickness of the material. For $h = 0$ we recover the local coupling $g(q)^2 = g_F^2$ model from the main text. The non-local bare electron-phonon interactions are $v_{ph}^{LO/TO}(q,\inu_n) = g(q)^2 2 \omega_e / [(\inu_n)^2 - \omega_e^2]$. In Fig.~\ref{fig:nonlocalPhonon} we show the leading eigenvalue $\lambda$ of the superconducting gap equation at $T=58$\,K (which is the critical temperature $T_c$ if Coulomb contributions are neglected), where $h=2$\,\AA\,and $g_F^2 = 0.7$\,eV$^2$. The phonon energy was set to $\omega_e = 0.3$\,eV, as in the main text.
We find that non-locality in the bare electron-phonon interaction does not qualitatively change the results discussed in the main text. This can be understood from the structure of the Fr\"ohlich interaction, which strongly suppresses the electron-phonon interaction in the $q \gtrsim 1/h \approx 1.3 k_F$ regime (and thus reduces $T_c$ by reducing the effective $\lambda_{ph}$), while the interplay between phonon and plasmon branches takes place at much smaller momenta $q \approx 0.05 k_F$ (see the main text). The interplay is thus only weakly affected by the non-locality of the electron-phonon interaction and as a result the qualitative structure of the plasmon- and phonon-mediated regimes is not changed.

\subsection{Non-Local Background Screening}

\begin{figure}
    \includegraphics[width=0.99\textwidth]{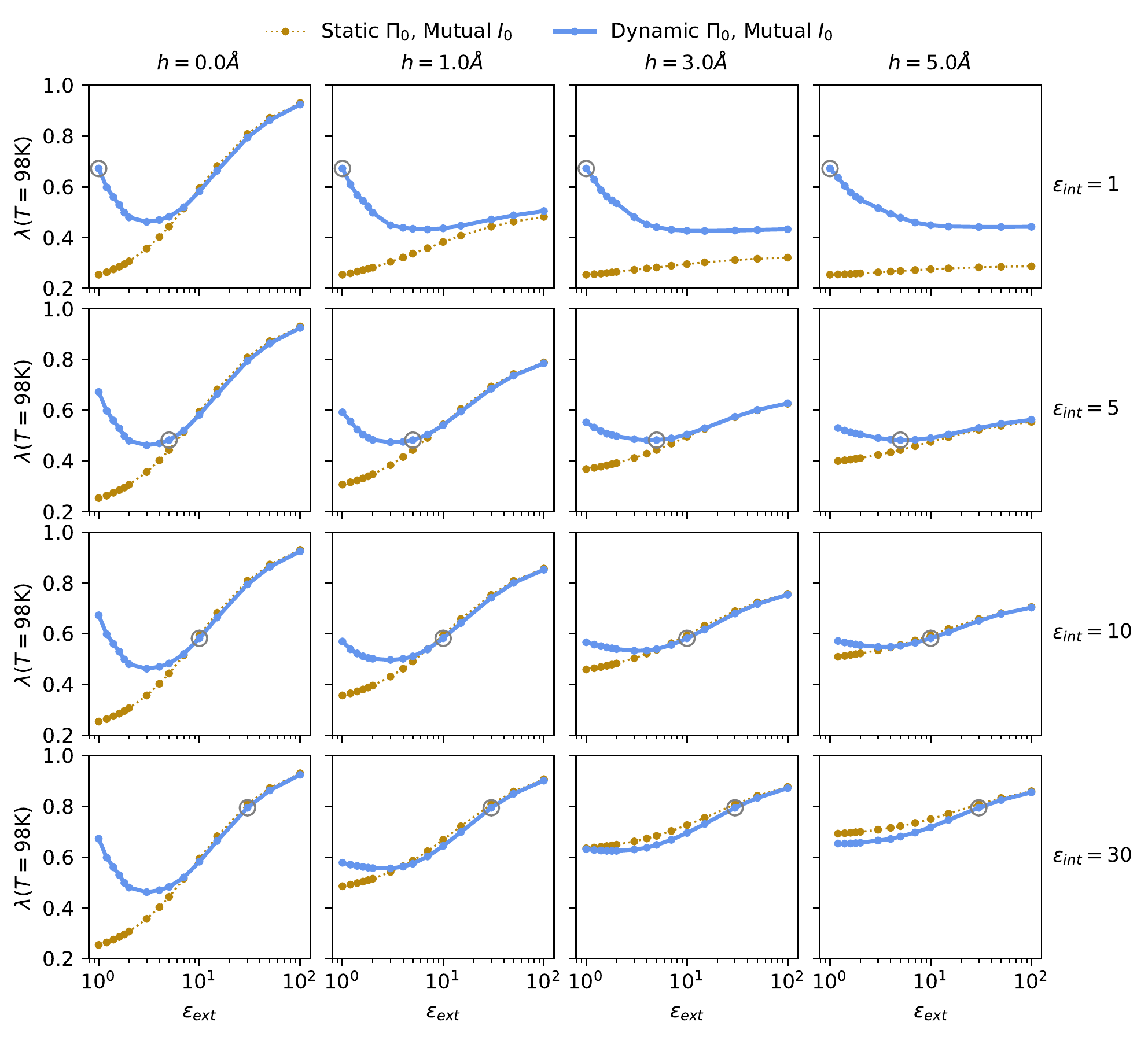}
    \caption{\label{fig:nonlocalScreeningEvs}
    The leading eigenvalue $\lambda$ at $T = 98$\,K including non-local background screening as a function of external screening $\varepsilon_{ext}$, for a variety of effective material thicknesses $h$ and internal screening $\varepsilon_{int}$.
    (Yellow-dotted) Statically mutually screened $I^{stat}_0(q, \inu_n)$.
    (Blue-solid) Fully dynamically and mutually screened $I_0(q, \inu_n)$. The gray circles surround the point where $\varepsilon_{ext} = \varepsilon_{int}$.
    }
\end{figure}

\begin{figure}
    \centering
    \includegraphics[width=0.6\textwidth]{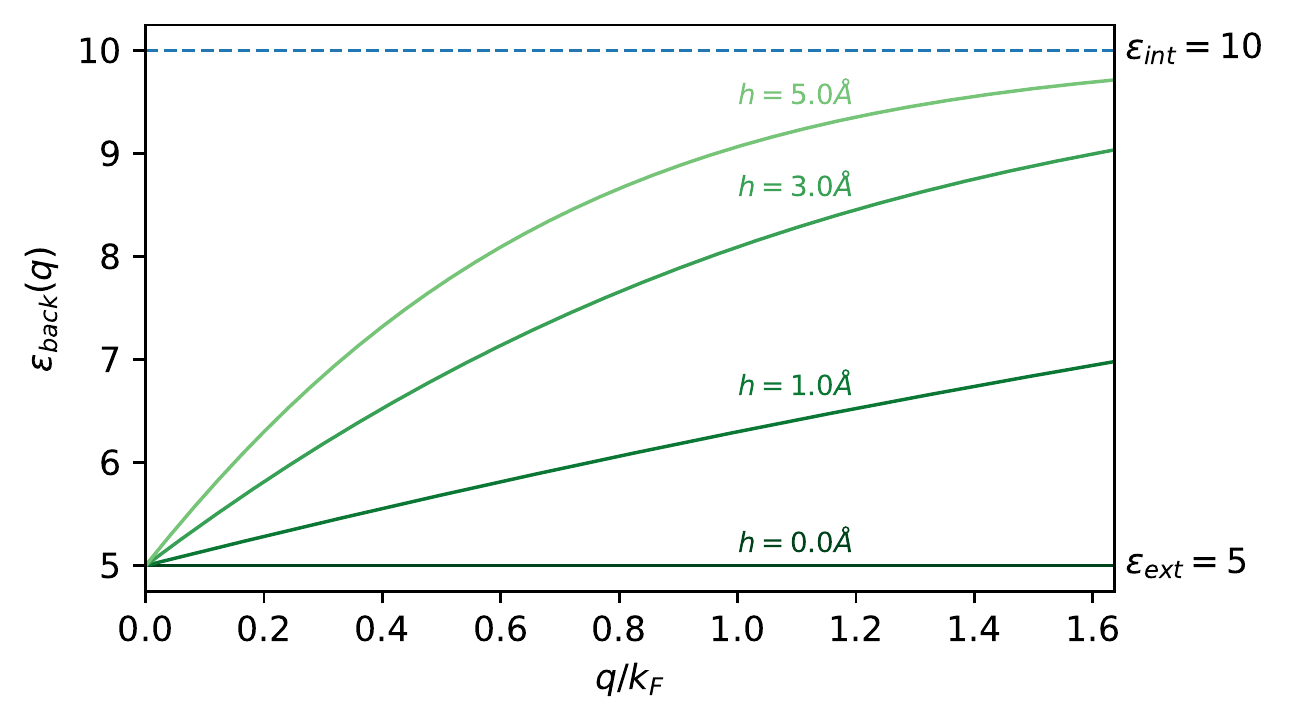}
    \caption{\label{fig:nonlocalScreeningEps}
    The background dielectric function $\varepsilon_{back}(q)$ for a variety of $h$, at $\varepsilon_{int} = 10$ and $\varepsilon_{ext} = 5$.
    }
\end{figure}

In Fig.~\ref{fig:nonlocalScreeningEvs} we show the superconducting leading eigenvalue $\lambda$ at $T = 98$\,K as a function of external screening $\varepsilon_{ext}$. We compare $\lambda$ including (blue-solid) and excluding (yellow-dotted) plasmonic contributions to understand the behaviour of the plasmon and phonon mediated regimes upon varying material properties $h$ and $\varepsilon_{int}$. Similar to the conclusions of the main text, we find that plasmonic contributions to superconductivity are strong when the total screening is weak. The effective height $h$ tunes whether $\varepsilon_{ext}$ or $\varepsilon_{int}$ contributes most to the total screening, as shown in Fig.~\ref{fig:nonlocalScreeningEps}. For example, at $h=0$ the background screening is completely determined by the external screening at all momenta $q$, such that plasmonic enhancement of $\lambda$ is found only when $\varepsilon_{ext} \lesssim 10$. On the other hand, for $h = 5$\,\AA\, the effect of $\varepsilon_{ext}$ is rather weak, and we only find plasmonic enhancement for weak internal screening $\varepsilon_{int} \lesssim 10$.

From this, we clearly understand that if the overall screening is weak plasmons can enhance superconductivity (as compared to the situation with electron-phonon and statically screening Coulomb interaction only), the interplay between enhanced electron-plasmon and electron-phonon interactions generically suppresses superconductivity as a result of the simultaneously present and equally strong static Coulomb repulsion in this regime, and that for enhanced overall screening the systems behaves as a conventional superconductor under the influence of static Coulomb repulsion.

\bibliography{bib}

\end{document}